\documentclass[twocolumn,floatfix,eqsecnum,aps]{revtex4-2}
\usepackage{graphicx}
\usepackage{amsmath}
\usepackage{amssymb}
\usepackage{bm}
\usepackage{hyperref}

\begin{document}

\title{Single-cone Dirac edge states on a lattice}
\author{A. Don\'{i}s Vela}
\affiliation{Instituut-Lorentz, Universiteit Leiden, P.O. Box 9506, 2300 RA Leiden, The Netherlands}
\author{C. W. J. Beenakker}
\affiliation{Instituut-Lorentz, Universiteit Leiden, P.O. Box 9506, 2300 RA Leiden, The Netherlands}

\date{November 2024}

\begin{abstract}
The stationary Dirac equation $(\bm{p}\cdot\bm{\sigma})\psi=E\psi$, confined to a two-dimensional (2D) region, supports states propagating along the boundary and decaying exponentially away from the boundary. These edge states appear on the 2D surface of a 3D topological insulator, where massless fermionic quasiparticles are governed by the Dirac equation and confined by a magnetic insulator. We show how the continuous system can be simulated on a 2D square lattice, without running into the fermion-doubling obstruction. For that purpose we adapt the existing tangent fermion discretization on an unbounded lattice to account for a lattice termination that simulates the magnetic insulator interface.
\end{abstract}
\maketitle

\section{Introduction}

The Dirac equation is a first order partial differential equation, which does not allow for the usual $\psi=0$ boundary condition of the Schr\"{o}dinger equation. Instead, boundary conditions relate the components of the Dirac spinor \cite{Alo97,Ben17}. For example, 
\begin{equation}
\lim_{x\rightarrow 0}[\psi_1(x,y)+ i\psi_2(x,y)]=0\label{infinitemassbc}
\end{equation}
represents an impenetrable boundary along the $y$-axis for the massless Dirac equation in two dimensions (2D),
\begin{equation}
-i\hbar v\begin{pmatrix}
0&\partial_x-i\partial_y\\
\partial_x+i\partial_y&0
\end{pmatrix}\begin{pmatrix}
\psi_1\\
\psi_2
\end{pmatrix}=E\begin{pmatrix}
\psi_1\\
\psi_2
\end{pmatrix}.
\end{equation}
Eq.\ \eqref{infinitemassbc} is known as the MIT bag boundary condition \cite{Joh75,Cho74,Arr19} or the infinite-mass boundary condition \cite{Ber87,Sto19}. An alternative boundary condition enforces one of the components of the spinor to vanish, say
\begin{equation}
\lim_{x\rightarrow 0}\psi_1(x,y)=0,\label{zigzagbc}
\end{equation}
while $\psi_2$ remains unconstrained. This applies to the zigzag edge in graphene \cite{Bre06,Akh08}.

In this paper we develop a method to implement these, and other, boundary conditions on a lattice, as a way to solve the Dirac equation numerically. Lattice-free numerical schemes exist \cite{Ant24}, the advantage of a lattice formulation is that it can be readily applied to disordered and interacting systems \cite{Zak24}. 

The physical system we have in mind is the 2D surface of a 3D topological insulator, where the spectrum contains a single Dirac cone \cite{Has10}. We are then faced with the following problem: The fermion doubling obstruction \cite{Nie81} prevents a local single-cone discretization of the Dirac equation ${\cal H}\psi=E\psi$ that preserves the fundamental symmetries (time reversal symmetry, $\sigma_y{\cal H}^\ast\sigma_y={\cal H}$, and chiral symmetry, $\sigma_z{\cal H}=-{\cal H}\sigma_z$). A variety of methods around the obstruction exist \cite{Tong}, we will focus on a symmetry-preserving approach that produces a local generalized eigenvalue equation \cite{Sta82,Pac21} (of the form ${\cal H}\psi=E{\cal P}\psi$ with local operators on both sides of the equation). This ``tangent fermion'' approach (the name refers to the replacement of the linear Dirac dispersion by a tangent) has so far been applied to systems with periodic boundary conditions \cite{Bee23}, here we extend the method to account for hard-wall boundaries.

\begin{figure}[tb]
\centerline{\includegraphics[width=0.8\linewidth]{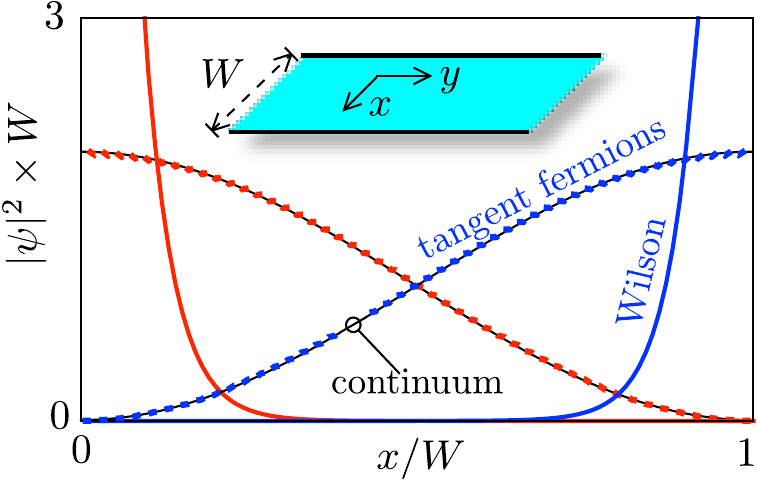}}
\caption{Colored curves: Wave function profiles of the discretized Dirac equation in a channel along the $y$-axis (width $W$, longitudinal momentum $k_y=50/W$) with infinite-mass boundary conditions ($\psi=\pm\sigma_y\psi$ at opposite boundaries). The red and blue curves represent the energy levels at $\pm E$ closest to $E=0$. The solid and dashed curves compare two alternative discretization schemes on a square lattice (lattice constant $a=W/100$), dashed for tangent fermions and solid for Wilson fermions (unit Wilson mass, see Sec.\ \ref{sec_Wilson}). The black solid curves are the solutions in the continuum [a cosine profile with an offset, at $E=(\hbar v/2W)\sqrt{\pi^2+(2k_yW)^2}$].
}
\label{fig_edgemode}
\end{figure}

Of particular interest in this context is the appearance of states that are confined to the boundary \cite{Bis22}. Such edge states exist for the zig-zag boundary condition \eqref{zigzagbc}, but not for the infinite-mass boundary condition \eqref{infinitemassbc}. Fig.\ \ref{fig_edgemode} shows that tangent fermions accurately follow the continuum solution, while the alternative Wilson fermion discretization \cite{Wil74} does not: a spurious edge state appears along infinite-mass boundaries for Wilson fermions \cite{Ara19}.

The outline of the paper is as follows. In the next section we show how a zero-current boundary condition can be incorporated into the generalized eigenvalue problem on a 2D square lattice. In Sec.\ \ref{sec_channel} we use this to study the edge states that appear for the zigzag boundary condition, and for more general boundary conditions that interpolate between zigzag and infinite mass. On the surface of a topological insulator, the entire class of boundary conditions is realized at the interface with a magnetic insulator, dependent on the orientation of the magnetization \cite{Bee24}. We compare with two alternative discretization schemes and with one lattice-free calculation in Secs.\ \ref{sec_comparison} and \ref{sec_lattice_free}. We then conclude in Sec.\ \ref{sec_conclude} with a discussion of the limitations of our approach.

\section{Tangent fermions in a confined geometry}

\subsection{Eigenvalue problem in the continuum}

In the continuum, we wish to solve the eigenvalue problem consisting of the 2D Dirac equation,
\begin{equation}
-i\hbar v(\sigma_x\partial_x+\sigma_y\partial_y)\psi(x,y)=E\psi(x,y),\label{HDirac}
\end{equation}
in a domain $(x,y)\in{\cal D}$,
with boundary condition
\begin{equation}
\psi(x,y)=(\bm{t}\cdot\bm{\sigma})\psi(x,y),\;\;(x,y)\in\delta{\cal D}.\label{eq_bc1}
\end{equation}
To ensure that no current flows through the boundary, the unit vector $\bm{t}=(t_x,t_y,t_z)$ should be orthogonal to the in-plane outward normal vector $\bm{n}=(n_x,n_y,0)$. 

For a specific example, if the boundary is the $y$-axis [outward normal $\bm{n}=(\pm 1,0,0)$] the zero-current boundary condition is
\begin{equation}
\psi(0,y)=(\sigma_y\cos\theta+\sigma_z\sin\theta)\psi(0,y),\label{eq_bc2}
\end{equation}
dependent on a single parameter $\theta$. 

Physically, on the surface of a 3D topological insulator, confinement of the massless Dirac fermions is enforced by the deposition of a magnetic insulator in the outside region $\bm{n}\cdot \bm{r}>0$. The magnetic energy $\bm{M}\cdot\bm{\sigma}$, with magnetization
\begin{equation}
\bm{M}=
\operatorname{sign}(n_x) M_0(0,-\sin\theta, \cos\theta),
\end{equation}
produces the boundary condition \eqref{eq_bc2} in the limit $M_0\rightarrow+\infty$. The case $\theta=0$ of a perpendicular magnetization is the \textit{infinite-mass} boundary condition. The case $\theta=\pi/2$ of a parallel magnetization corresponds to the \textit{zigzag} boundary condition in graphene, we will use that name even though we are not considering the graphene lattice.

\subsection{Generalized eigenvalue problem on the lattice}

We discretize the system on a 2D square lattice, lattice constant $a$. The local finite difference discretization of the derivative,
\begin{equation}
\frac{df}{dx}\mapsto\frac{1}{2a}[f(x+a)-f(x-a)],
\end{equation}
produces a sine dispersion $E(k)=(\hbar v/a) \sin ak$ with a spurious second Dirac cone at the boundary $k=\pi/a$ of the Brillouin zone. As shown by Stacey \cite{Sta82}, this fermion doubling can be avoided at the expense of a nonlocal discretization,
\begin{equation}
\frac{df}{dx}\mapsto \frac{2}{a}\sum_{n=1}^\infty(-1)^n[f(x-na)-f(x+na)],
\end{equation}
with tangent dispersion
\begin{equation}
E(k)=\pm(2\hbar v/a)\tan(ak/2).
\end{equation}

Following Ref.\ \onlinecite{Pac21}, we implement Stacey's tangent fermions via a \textit{local} generalized eigenvalue problem,
\begin{subequations}
\label{HPdef}
\begin{align}
{\cal H}\Psi={}&E{\cal P}\Psi,\\
{\cal P}={}&\tfrac{1}{4}(1+\cos a\hat{k}_x)(1+\cos a\hat{k}_y),\\
{\cal H}={}&\frac{\hbar v}{2a}\bigl[\sigma_x(1+\cos a\hat{k}_y)\sin a\hat{k}_x\nonumber\\
&+\sigma_y(1+\cos a\hat{k}_x)\sin a\hat{k}_y\bigr],
\end{align}
\end{subequations}
in terms of the displacement operators
\begin{equation}
\begin{split}
&(\cos a\hat{k}_x) f(x,y)=\tfrac{1}{2}[f(x+a,y)+f(x-a,y)],\\
&(\sin a\hat{k}_x) f(x,y)=\tfrac{1}{2i}[f(x+a,y)-f(x-a,y)],
\end{split}
\end{equation}
and similarly for $\cos a\hat{k}_y$ and $\sin a\hat{k}_y$. 

The tangent fermion discretization \eqref{HPdef} preserves the fundamental symmetries of the continuum Dirac equation \eqref{HDirac}: time-reversal symmetry (${\cal H}$ and ${\cal P}$ are invariant if both $\bm{k}$ and $\bm{\sigma}$ change sign) and chiral symmetry (${\cal H}$ anticommutes with $\sigma_z$). The 2D dispersion relation is
\begin{equation}
E=\pm\frac{2\hbar v}{a}\sqrt{\tan^2(ak_x/2)+\tan^2(ak_y/2)}.
\end{equation}

Both operators ${\cal H}$ and ${\cal P}$ in Eq.\ \eqref{HPdef} are Hermitian and ${\cal P}$ is also positive definite, These two properties are both needed to ensure that ${\cal H}\Psi=E{\cal P}\Psi$ has \textit{real} eigenvalues $E$. Moreover, ${\cal H}$ and ${\cal P}$ are sparse matrices, only nearby sites on the lattice are coupled, which allows for an efficient calculation of the energy spectrum.

\subsection{Implementation of the boundary condition on the lattice}
\label{tangentbc}

Any modification of the generalized eigenvalue problem \eqref{HPdef} to account for the boundary condition must preserve the Hermiticity of ${\cal H}$ and ${\cal P}$ as well as the positive definiteness of ${\cal P}$. Allowed operations include unitary transformations and the removal of a set of rows and columns with the same index (because any principal submatrix of a positive definite matrix remains positive definite). 

Starting from the infinite lattice, we first remove the rows and columns of ${\cal H}$ and ${\cal P}$ that refer to sites outside of the system. If the system contains $N$ sites the dimension of the resulting principal submatrices is $2N\times 2N$, including the spin degree of freedom on each site. We denote by $N_B$ the number of sites on the boundary and by ${\cal B}$ the set of these lattice points.

To each of the $N_B$ boundary points  we associate a unit vector $\bm{t}_n$. The $2\times 2 $ unitary matrix $U_n$ rotates $\bm{t}_n\cdot\bm{\sigma}$ to the $\sigma_z$ Pauli matrix,
\begin{equation}
U_n(\bm{t}_n\cdot\bm{\sigma})U_n^\dagger=\sigma_z.
\end{equation}
We construct a $2N\times 2N$ block-diagonal unitary matrix
\begin{equation}
{\cal U}_{nm}=\delta_{nm}\times\begin{cases}
U_n&\text{if}\;\;n\in{\cal B},\\
\sigma_0&\text{if}\;\;n\notin{\cal B},
\end{cases}
\end{equation}
with $\sigma_0$ the $2\times 2$ unit matrix, and perform the unitary transformations
\begin{equation}
{\cal H}\mapsto{\cal U}^\dagger {\cal H}{\cal U},\;\;{\cal P}\mapsto{\cal U}^\dagger {\cal P}{\cal U}.
\end{equation}

The transformed boundary condition ${\psi}=\sigma_z{\psi}$ expresses that the spin-down component of ${\psi}$ vanishes. We implement that on the lattice by removing from the matrices ${\cal U}^\dagger {\cal H}{\cal U}$ and ${\cal U}^\dagger {\cal P}{\cal U}$ the spin-down row and column on each site $n\in{\cal B}$. The resulting matrices $\tilde{\cal H}$ and $\tilde{\cal P}$ have dimension $(2N-N_B)\times (2N-N_B)$. We thus arrive at the boundary-constrained generalized eigenvalue problem
\begin{equation}
\tilde{\cal H}{\psi}=E\tilde{\cal P}{\psi}.\label{tildeHP}
\end{equation}

\section{Edge states in a channel geometry}
\label{sec_channel}

We test the validity of the boundary-constrained generalized eigenvalue problem \eqref{tildeHP} in a channel geometry. The channel is aligned with the lattice vectors, with boundaries at $x=0$ and $x=W$ and outward normal vectors $\bm{n}=(\pm 1,0,0)$. We allow for different boundary conditions at opposite edges,
\begin{equation}
\begin{split}
\psi(0,y)=(\sigma_y\cos\theta_1+\sigma_z\sin\theta_1)\psi(0,y),\\
\psi(W,y)=(\sigma_y\cos\theta_2+\sigma_z\sin\theta_2)\psi(W,y).
\end{split}
\label{bctheta}
\end{equation}
The parallel momentum $k_y\equiv q$ is a good quantum number, so to compute the dispersion relation $E(q)$ we can work on a 1D lattice (sites $n=1,2,\ldots N\equiv W/a$, Brillouin zone $-\pi/a<q<\pi/a$).

\begin{figure*}[tb]
\centerline{\includegraphics[width=1\linewidth]{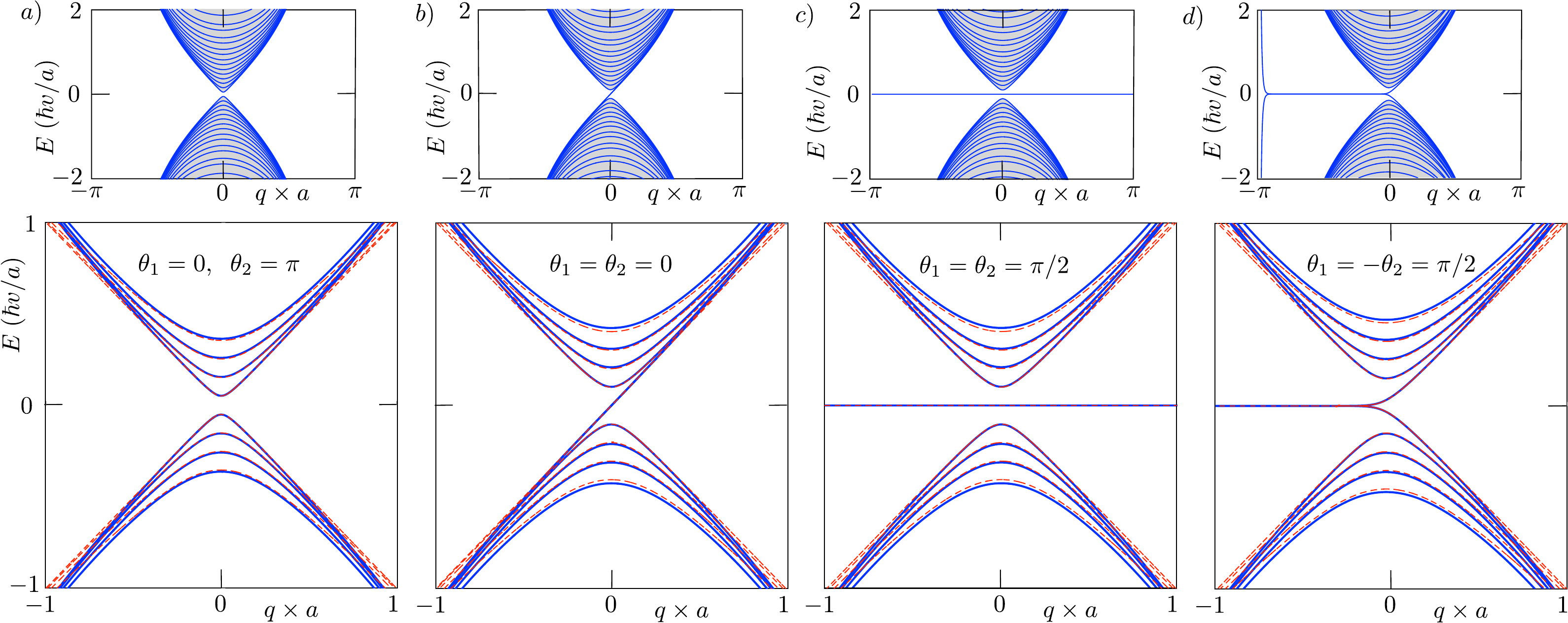}}
\caption{Blue solid curves: Dispersion relation of tangent fermions confined to a channel along the $y$-axis (width $W=31a$), computed from the boundary-constrained generalized eigenvalue problem \eqref{tildeHP}. The top panels show the full Brillouin zone, the bottom panels show the low-energy modes on a larger scale. In the bottom panels we have included the continuum limit \eqref{Echannelq} (red dashed). Blue and red curves are barely distinguishable for the lowest modes. In the top panels the Dirac cone of bulk states is shaded grey, states outside of the cone are edge states.
}
\label{fig_masszigzag}
\end{figure*}

\begin{figure}[tb]
\centerline{\includegraphics[width=1\linewidth]{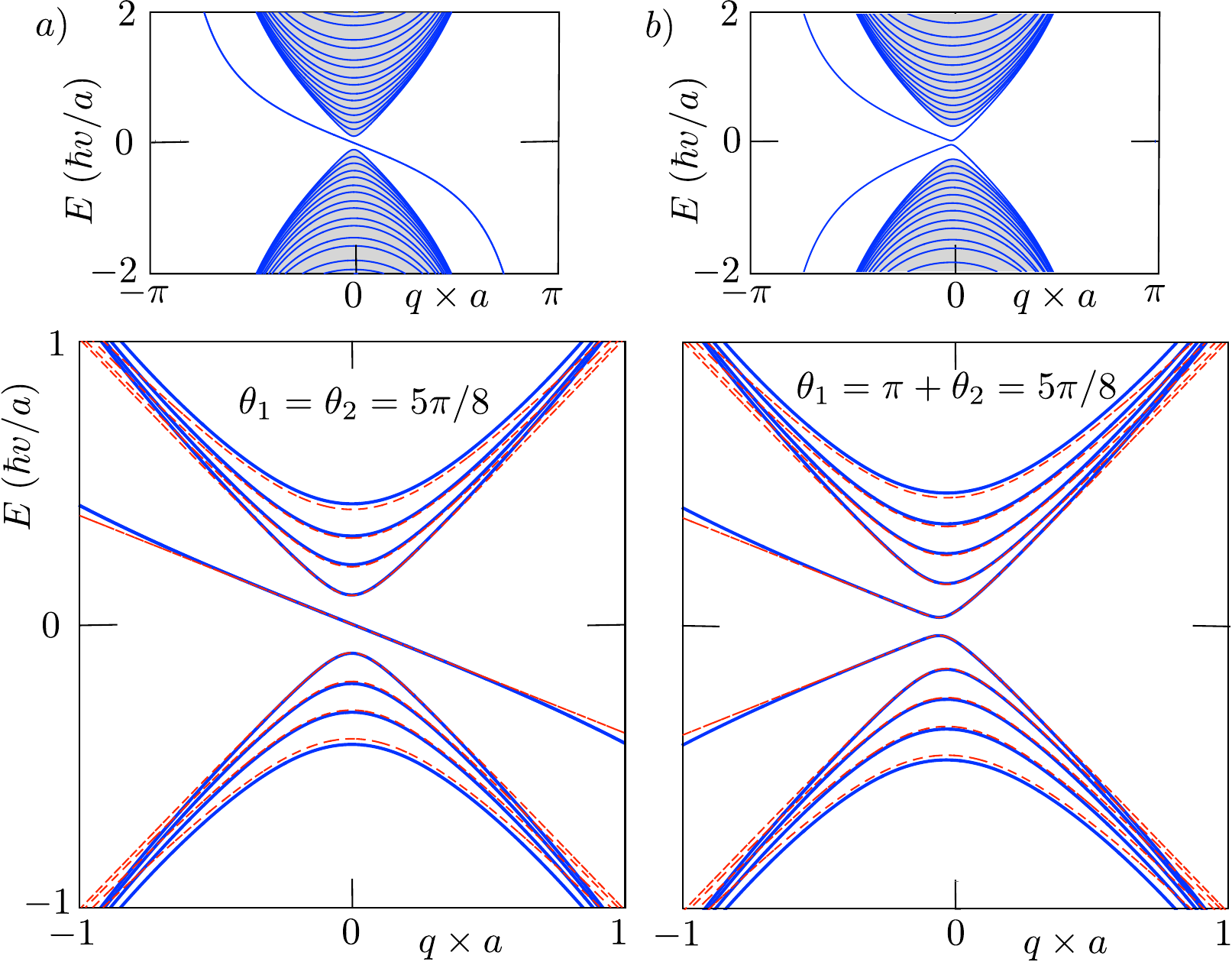}}
\caption{Same as Fig.\ \ref{fig_masszigzag} for boundary conditions intermediate between the infinite mass and zigzag cases. Panels \ref{fig_other}a and \ref{fig_other}b approach panels \ref{fig_masszigzag}c and \ref{fig_masszigzag}d, respectively, as the angle $5\pi/8$ is rotated to $\pi/2$.
}
\label{fig_other}
\end{figure}

The two unitaries that rotate the boundary condition at sites $n=1$ and $n=N$ into $\psi=\sigma_z\psi$ are
\begin{equation}
U_1=e^{i(\theta_1/2-\pi/4)\sigma_x},\;\;U_N=e^{i(\theta_2/2-\pi/4)\sigma_x}.
\end{equation}
Results for various choices of $\theta_1,\theta_2$ are shown in Figs.\ \ref{fig_masszigzag} and \ref{fig_other}. Panels a) and b) in Fig.\ \ref{fig_masszigzag} show the case of an infinite-mass boundary condition, when there is no edge state \cite{Cho74,Joh75,Arr19,Ber87}. Panels c) and d) show the case of a zigzag boundary condition \cite{note1}, when there is a dispersionless edge state \cite{Fuj96,Wak01,Koh07}. Fig.\ \ref{fig_other} shows boundary conditions intermediate between the infinite mass and the zigzag case, when the edge state acquires a dispersion \cite{Vol09,Tka09,Ost11}.

In all cases the low-energy modes on the lattice agree very well with the dispersion relation in the continuum \cite{Bee24},
\begin{subequations}
\label{Echannelq}
\begin{align}
& E\cos\alpha_- =\frac{k(E)\sin\alpha_-}{\tan[k(E)W]} + q\cos\alpha_+,\\
&k(E)=\sqrt{E^2-q^2},\;\;\alpha_\pm=\tfrac{1}{2}(\theta_1\pm\theta_2).
\end{align}
\end{subequations}
The agreement is reached without any adjustable parameter.

\section{Comparison with other single-cone discretizations}
\label{sec_comparison}

\subsection{Wilson fermions}
\label{sec_Wilson}

The Wilson approach \cite{Wil74} to avoid fermion doubling is to add a momentum dependent term $\propto\sigma_z$ to the discretized Dirac Hamiltonian that gaps the doublers at the edge of the Brillouin zone,
\begin{subequations}
\label{HWilsondef}
\begin{align}
&{\cal H}_{\rm Wilson}\Psi=E\Psi,\\
&{\cal H}_{\rm Wilson}=\frac{\hbar v}{a}\bigl[\sigma_x\sin a\hat{k}_x+\sigma_y\sin a\hat{k}_y\bigr]\nonumber\\
&\qquad+\frac{\hbar v}{a}M_{\rm Wilson}\sigma_z(2-\cos a\hat{k}_x-\cos a\hat{k}_y).
\end{align}
\end{subequations}
The term $\propto M_{\rm Wilson}\sigma_z$ breaks time-reversal symmetry (${\cal H}_{\rm Wilson}$ is not invariant if both $\bm{k}$ and $\bm{\sigma}$ change sign) and it breaks chiral symmetry $({\cal H}_{\rm Wilson}$ does not anticommute with $\sigma_z$).

We implement the boundary condition in the same way as we did in Sec.\ \ref{tangentbc} for tangent fermions, but now we have a conventional eigenvalue problem rather than a generalized eigenvalue problem, so there is only one operator to transform:
\begin{equation}
\tilde{\cal H}_{\rm Wilson}\psi=E\psi,\label{calHWilson}
\end{equation}
where $\tilde{\cal H}_{\rm Wilson}$ is obtained from the unitary transformation ${\cal U}^\dagger{\cal H}_{\rm Wilson}{\cal U}$ upon removal of the spin-down row and column from each boundary site.

\begin{figure}[tb]
\centerline{\includegraphics[width=1\linewidth]{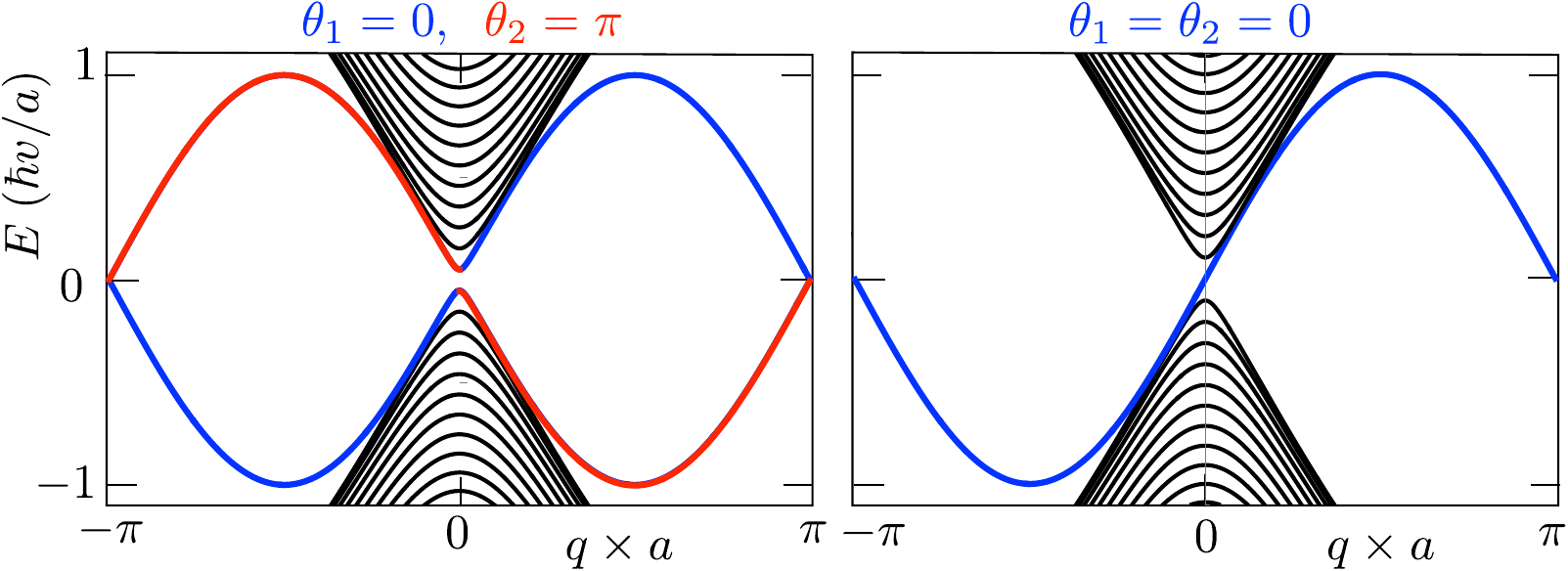}}
\caption{Dispersion relation of Wilson fermions confined to a channel with infinite-mass boundary conditions (channel width $W = 31 a$, Wilson mass $M_{\rm Wilson}=1$). The color of the edge modes indicates to which edge they are bound (blue on edge 1, red on edge 2 in the left panel; blue on both edges in the right panel).
}
\label{fig_wilson}
\end{figure}

In Fig.\ \ref{fig_wilson} we show the spectrum for the case of an infinite-mass boundary condition. In contrast to the tangent fermion spectrum (Fig.\ \ref{fig_masszigzag}a,b), there is now a sinusoidally dispersing edge mode. The appearance of an edge mode for Wilson fermions was noticed previously \cite{Ara19}. It is a lattice artefact in the sense that the Dirac equation in the continuum has no edge mode for an infinite-mass boundary condition.

The Wilson fermion edge state dispersion in Fig.\ \ref{fig_wilson} connects the Brillouin zone boundary at $q=\pm\pi/a$ to the bulk states around $q=0$. The edge state is an eigenstate of $\sigma_y$, with eigenvalue $\pm 1$ and dispersion $E(q)=\pm(\hbar v/a)\sin qa$ dictated by the boundary condition $\psi=\pm\sigma_y\psi$ at each edge. The state is bound to the edge on a length scale $\simeq a/M_{\rm Wilson}$, it merges with the bulk bands in the limit $M_{\rm Wilson}\rightarrow 0$.

A calculation for a single boundary at $x=0$, boundary condition $\psi(0)=\sigma_y\psi(0)$, gives the wave function of the Wilson fermion edge state at energy $E=(\hbar v/a)\sin qa$,
\begin{equation}
\begin{split}
&\psi(x)=e^{-\kappa x/a}\sqrt{\frac{\kappa}{a}}{1\choose i},\;\;x>0,\\
&\kappa=|M_{\rm Wilson}|\bigl[2-\cos qa-\operatorname{sign}(M_{\rm Wilson})\bigr].
\end{split}
\end{equation}
Near $q=\pi/a$ the decay length $a/\kappa$ is $a/2M_{\rm Wilson}$ or $-a/4M_{\rm Wilson}$, respectively, for positive or negative Wilson mass.

\subsection{Staggered fermions}

On a 1D lattice the single-cone dispersion
\begin{equation}
E(k)=\pm(2\hbar v/a)\sin(ak/2)
\end{equation}
can be obtained by applying a different lattice to each of the two spinor components. The two lattices are staggered, displaced by half a lattice constant. This staggered fermion method to avoid fermion doubling, due to Susskind \cite{Sus77}, is not effective in 2D: The staggered 2D lattice has a pair of inequivalent Dirac points in the Brillouin zone, at the center and at the corner \cite{Bee24}.

Hern\'{a}ndez and Lewenkopf \cite{Her12} worked around this obstruction by combining the staggered fermion $\sin(ak_x/2)$ dispersion in the $x$-direction with the tangent fermion $\tan(ak_y/2)$ dispersion in the $y$-direction. Their Hamiltonian has the form
\begin{align}
{\cal H}_{\rm staggered} ={}& [\sigma_x \sin a\hat{k}_x + \sigma_y (\cos a\hat{k}_x + 1)]\tan(a\hat{k}_y/2)\nonumber\\
& + \sigma_x \sin a\hat{k}_x+ \sigma_y (\cos a\hat{k}_x - 1),\label{Hstaggered}
\end{align}
with dispersion
\begin{equation}
E=\pm\frac{2\hbar v}{a}\frac{ \sqrt{1-\cos ak_x \cos ak_y}}{\sqrt{1+\cos ak_y}}.
\end{equation}
The Hamiltonian \eqref{Hstaggered} preserves chiral symmetry but breaks time-revesal symmetry.

To implement the boundary condition we proceed as we did for Wilson fermions,
\begin{equation}
\tilde{\cal H}_{\rm staggered}\psi=E\psi,\label{Htildestaggered}
\end{equation}
where $\tilde{\cal H}_{\rm staggered}$ is obtained from the unitary transformation ${\cal U}^\dagger{\cal H}_{\rm staggered}{\cal U}$ upon removal of the spin-down row and column from each boundary site.

Ref.\ \onlinecite{Her12} considered the case of zigzag boundary conditions, and we reproduce their spectrum, which is nearly the same as for tangent fermions (Figs.\ \ref{fig_masszigzag}c,d). The zigzag boundary condition preserves chiral symmetry. 

Once we break chiral symmetry, the agreement of staggered fermions with the continuum is lost. In particular, Eq.\ \eqref{Htildestaggered} produces a spurious edge mode for the infinite-mass boundary condition. We see no fundamental reason for this deficiency, it may well be possible to modify the approach so that it is no longer restricted to zigzag boundary conditions.

\section{Comparison with a lattice-free calculation}
\label{sec_lattice_free}

A lattice-free numerical solution of the Dirac equation in a confined geometry is developed in Ref.\ \onlinecite{Ant24}. That work uses infinite-mass boundary conditions, with a nonzero mass $m$ of the Dirac fermions inside the system. Although in the main text we took $m=0$, we can readily include a mass in the  tangent fermion discretization \cite{Pac21}: This enters as a term $m\sigma_z{\cal P}\Psi$ on the left-hand-side of  Eq.\ \eqref{HPdef}.

\begin{figure}[tb]
\centerline{\includegraphics[width=0.9\linewidth]{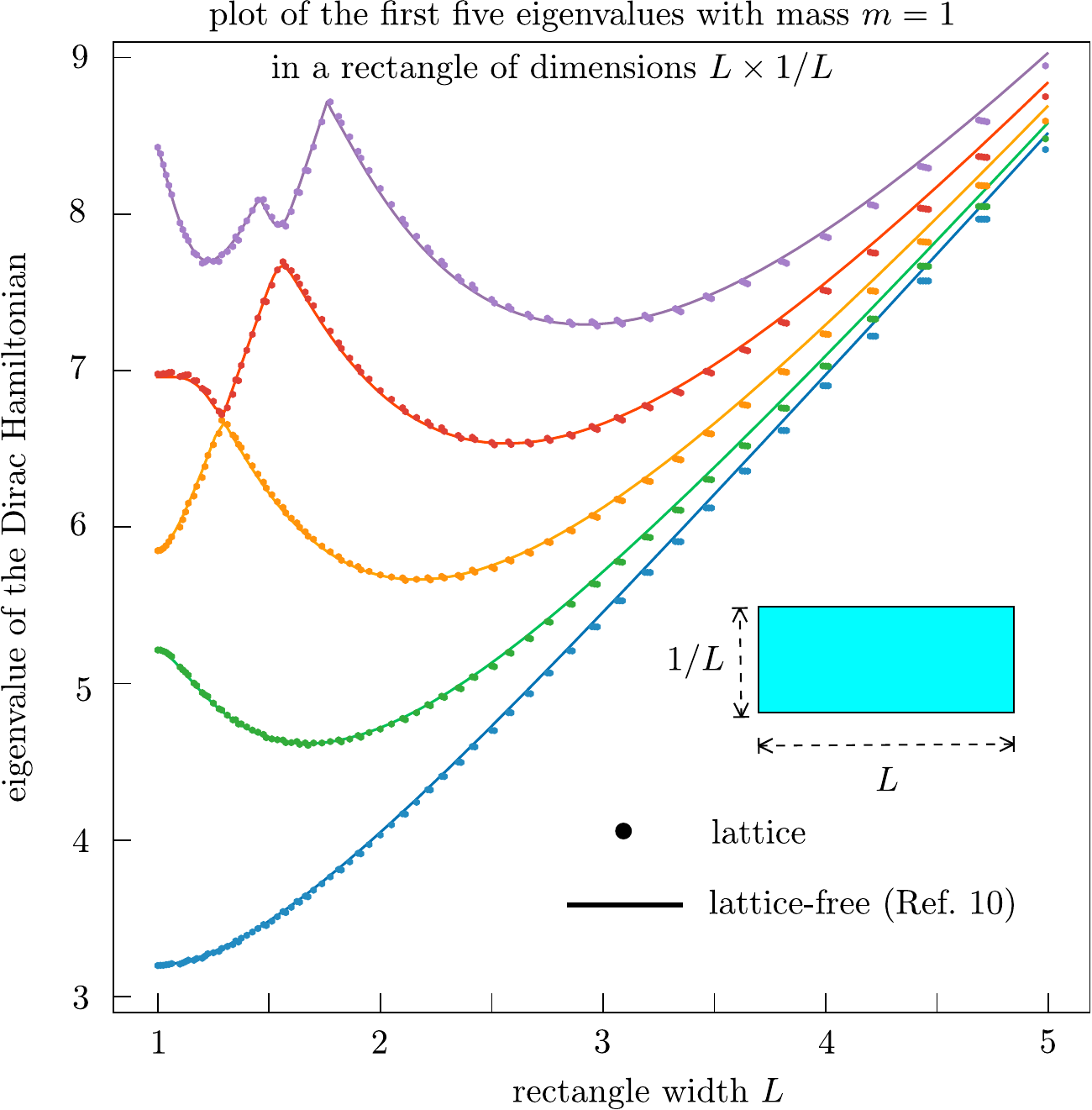}}
\caption{First five positive eigenvalues of the dimensionless Dirac equation, $-i\sigma_x\partial\psi/\partial x-i\sigma_y\partial\psi/\partial y+m\sigma_z\psi=E\psi$, in a rectangle of unit area with infinite-mass boundary conditions. The mass $m=1$ inside the rectangle, the aspect ratio of the rectangle is varied. The plot compares our lattice calculation (data points, lattice constant $1/80$) with the lattice-free calculation of Ref.\ \onlinecite{Ant24} (solid lines).
}
\label{fig_widths}
\end{figure}

\begin{figure}[tb]
\centerline{\includegraphics[width=0.9\linewidth]{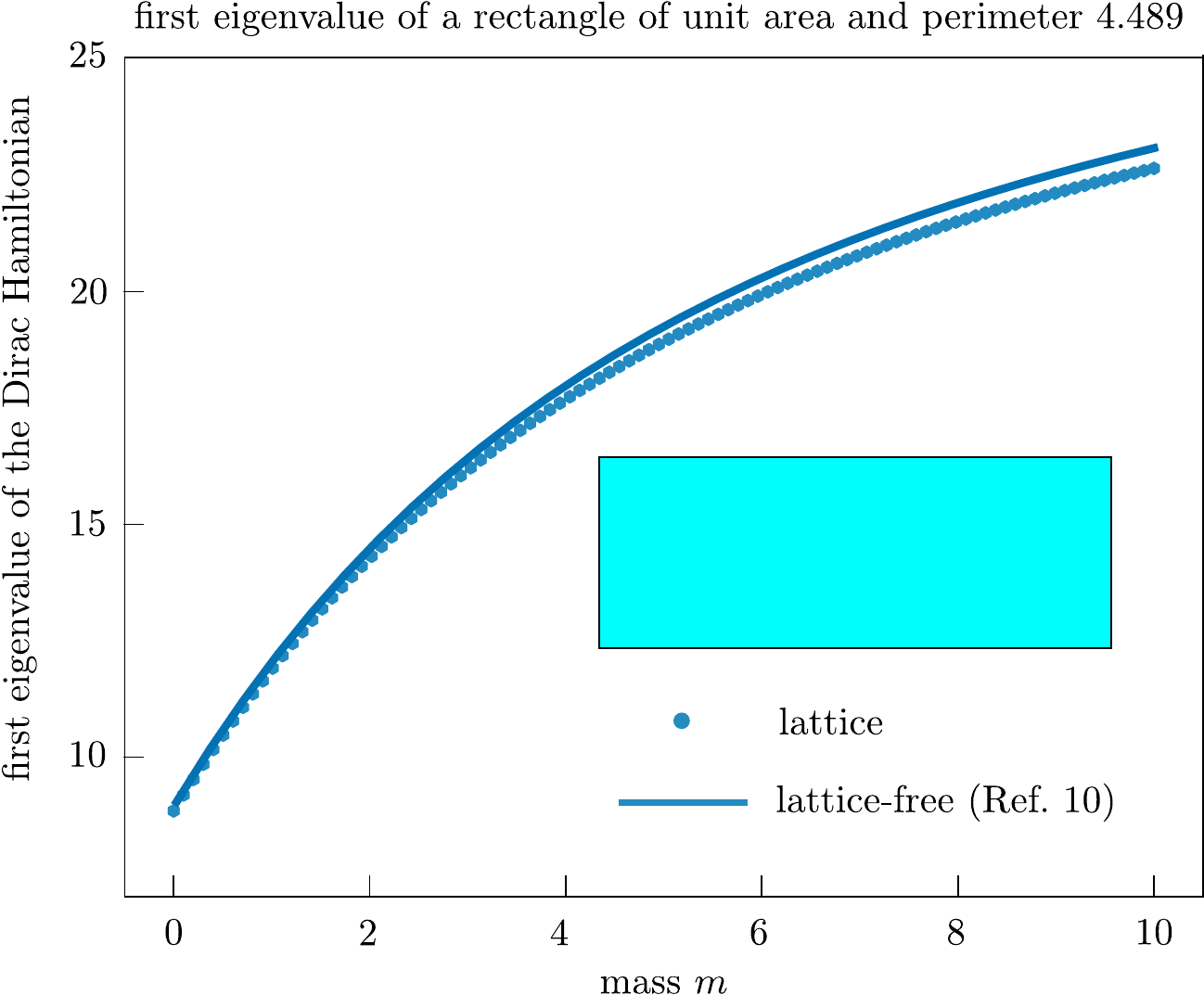}}
\caption{Same as Fig.\ \ref{fig_widths}, but now showing only the first eigenvalue as a function of the mass $m$, at fixed aspect ratio of the rectangle. For this plot the lattice constant is set at $1/68$.
}
\label{fig_mass}
\end{figure}

We compare results for a rectangle in Figs.\ \ref{fig_widths} and \ref{fig_mass}. The agreement is quite satisfactory.

\section{Conclusion}
\label{sec_conclude}

\begin{figure}[tb]
\centerline{\includegraphics[width=0.8\linewidth]{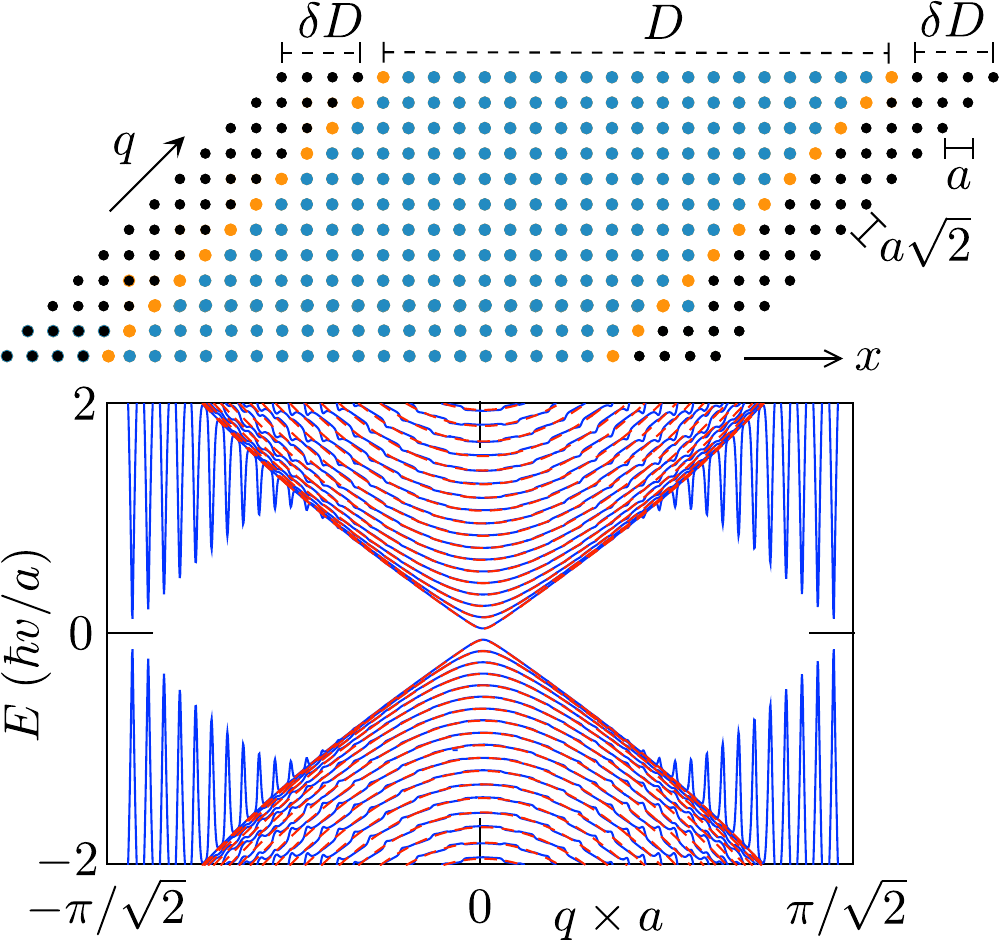}}
\caption{\textit{Blue solid curves}: Tangent-fermion band structure of a channel rotated by $45^\circ$ relative to the lattice vectors. The channel has width $W=Da/\sqrt 2$, with $D=45$ the number of sites along the $x$-axis. Infinite-mass boundary conditions are imposed at site number 1 and site number $D$ (orange dots). \textit{Red dashed curves}: The same band structure for a finite-mass confinement. A layer of $\delta D=15$ sites is added on both sides of the channel (black dots) and in this layer a mass term $M\sigma_z$ is introduced in the Dirac equation ($M=3\hbar v/a$). The two mass layers are connected by periodic boundary conditions on site number 1 and site number $D+2\delta D$.
}
\label{fig_oblique}
\end{figure}

In summary, we have shown how single-Dirac-cone boundary conditions can be implemented on a 2D lattice. The tangent fermion approach correctly reproduces the absence of edge states for the infinite-mass boundary condition and the dispersionless edge states for the zigzag boundary condition, see Fig.\ \ref{fig_masszigzag}.

The method does have its limitations. We mention two. Very good agreement with the continuum is obtained if the boundaries are aligned with the lattice vectors, but not for misaligned boundaries. In Fig.\ \ref{fig_oblique} we show the spectrum for a channel rotated by $45^\circ$. The periodicity along the channel is then increased by a factor $\sqrt{2}$, and the Brillouin zone is reduced by that factor. The infinite-mass boundary condition then introduces spurious oscillations in the band structure (blue curves). These can be avoided by working with a finite-mass boundary layer (red curves). The way we understand this lattice artefact, is that the pole in the tangent dispersion is folded into the reduced Brillouin zone and coupled to low-energy excitations by the infinite-mass boundary condition. A finite mass $M$ is transparent at energies $E>M$, leaving the pole uncoupled.

\begin{figure}[tb]
\centerline{\includegraphics[width=1\linewidth]{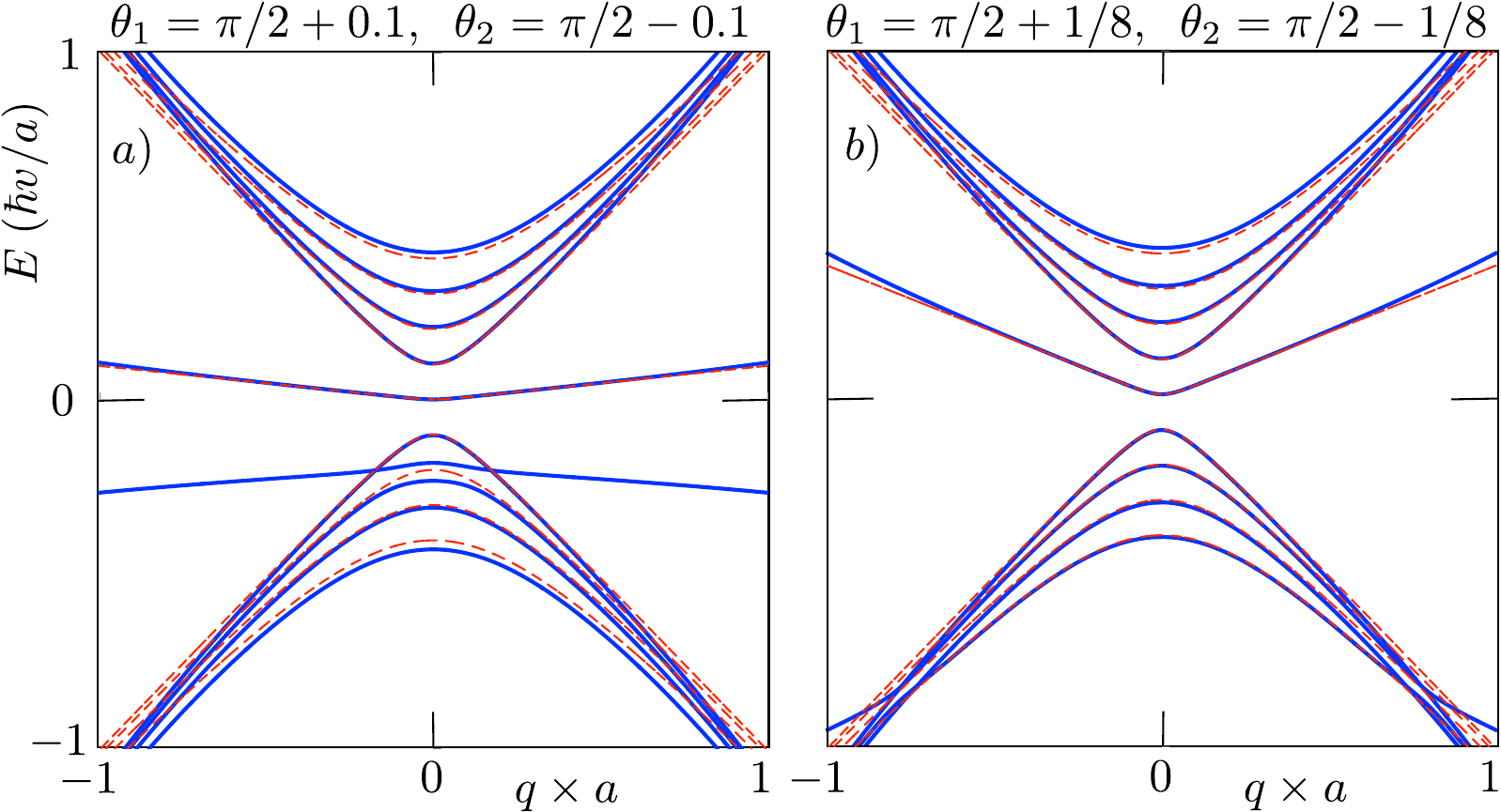}}
\caption{Panel a): Same as Fig.\ \ref{fig_masszigzag}c, for a small perturbation around the zigzag boundary condition $\theta_1=\theta_2=\pi/2$, in order to reveal the degeneracy doubling of the flat band. Only the upper band agrees with the continuum, the lower band is a lattice artefact. In panel b) the spurious lower band is pushed out of the low-energy range by moving a bit further away from the zigzag case.
}
\label{fig_splitzigzag}
\end{figure}

A second limitation is more fundamental: The flat band in Fig.\ \ref{fig_masszigzag}c is topologically equivalent to a zeroth Landau level, and therefore suffers from the doubled degeneracy problem of Ref.\ \onlinecite{Don23}. This degeneracy doubling is revealed if we slightly perturb the boundary condition, see Fig.\ \ref{fig_splitzigzag}a. The flat band splits into an upper band that agrees with the continuum result, and a spurious lower band. The effect of the spurious band can be minimized by pushing it away from $E=0$, see Fig.\ \ref{fig_splitzigzag}b.

\acknowledgments

Our tangent fermion project has benefited much from the insights of M. J. Pacholski. Discussions with A. R. Akhmerov are gratefully acknowledged. Financial support was received from the European Research Council (ERC),  under the European Union's Horizon 2020 research and innovation programme.

The computer codes used for Figs.\ \ref{fig_widths} and \ref{fig_mass} are available at \url{https://dx.doi.org/10.5281/zenodo.14883054}


\begin{thebibliography}{99}
\bibitem{Alo97} V. Alonso, S. de Vincenzo, and L. Mondino, \textit{On the boundary conditions for the
Dirac equation}, Eur. J. Phys. \textbf{18}, 315 (1997) .
\bibitem{Ben17} R. D. Benguria, S. Fournais, E. Stockmeyer, and H. Van Den Bosch, \textit{Self-adjointness of two-dimensional Dirac operators on domains}, Annales Henri Poincar\'{e} \textbf{18}, 1371 (2017).
\bibitem{Joh75} K. Johnson, \textit{The M.I.T. bag model}, Acta Phys. Polon. B \textbf{6}, 865 (1975).
\bibitem{Cho74} A. Chodos, R. L. Jaffe, K. Johnson, C. B. Thorn, and V. F. Weisskopf, \textit{New extended model of hadrons}, Phys. Rev. D \textbf{9}, 3471 (1974).
\bibitem{Arr19} N. Arrizibalaga, L. Le Treust, A. Mas, and N. Raymond, \textit{The MIT bag model as an infinite mass limit}, J.
Ec. Polytech. Math. \textbf{6}, 329 (2019).
\bibitem{Ber87} M. V. Berry and R. J. Mondragon, \textit{Neutrino billiards: time-reversal symmetry-breaking without magnetic fields}, Proc. R. Soc. Lond. A \textbf{412}, 53 (1987).
\bibitem{Sto19} E. Stockmeyer and S. Vugalter, \textit{Infinite mass boundary conditions for Dirac operators}, J. Spectr. Theory \textbf{9} 569 (2019).
\bibitem{Bre06} L. Brey and H.A. Fertig, \textit{Electronic states of graphene nanoribbons}, Phys. Rev. B \textbf{73}, 235411 (2006).
\bibitem{Akh08} A. R. Akhmerov and C. W. J. Beenakker, \textit{Boundary conditions for Dirac fermions on a terminated honeycomb lattice},  Phys. Rev. B \textbf{77}, 085423 (2008).
\bibitem{Ant24} P. R. S. Antunes, F. Bento, and D. Krej\v{c}i\v{r}\'{\i}k, \textit{Numerical optimisation of Dirac eigenvalues}, J. Phys. A \textbf{57}, 475203 (2024).
\bibitem{Zak24} V. A. Zakharov, J. Tworzyd{\l}o, C. W. J. Beenakker, and M. J. Pacholski, \textit{Helical Luttinger liquid on a space-time lattice}, Phys. Rev. Lett. \textbf{133}, 116501 (2024).
\bibitem{Has10} M. Z. Hasan and C. L. Kane, \textit{Topological insulators}, Rev. Mod. Phys. \textbf{82}, 3045 (2010).
\bibitem{Nie81} H. B. Nielsen and M. Ninomiya, \textit{A no-go theorem for regularizing chiral fermions}, Phys. Lett. B \textbf{105}, 219 (1981).
\bibitem{Tong} An overview of methods to avoid fermion doubling in lattice gauge theory can be found in chapter 4 of David Tong's lecture notes: \url{https://www.damtp.cam.ac.uk/user/tong/gaugetheory.html}.
\bibitem{Sta82} R. Stacey, \textit{Eliminating lattice fermion doubling}, Phys. Rev. D \textbf{26}, 468 (1982).
\bibitem{Pac21} M. J. Pacholski, G. Lemut, J. Tworzyd{\l}o, and C. W. J. Beenakker, \textit{Generalized eigenproblem without fermion doubling for Dirac fermions on a lattice}, SciPost Phys. \textbf{11}, 105 (2021).
\bibitem{Bee23} C. W. J. Beenakker, A. Don\'{i}s Vela, G. Lemut, M. J. Pacholski, and J. Tworzyd{\l}o, \textit{Tangent fermions: Dirac or Majorana fermions on a lattice without fermion doubling}, Annalen der Physik \textbf{535}, 2300081 (2023).
\bibitem{Bis22} S. Biswas and G. W. Semenoff, \textit{Massless fermions on a half-space: the curious case of 2+1-dimensions}, JHEP \textbf{10}, 045 (2022).
\bibitem{Wil74} K. G. Wilson, \textit{Confinement of quarks}, Phys. Rev. D \textbf{10}, 2445 (1974).
\bibitem{Ara19} A. L. Ara\'{u}jo, R. P. Maciel, R. G. F. Dornelas, D. Varjas, and G. J. Ferreira, \textit{Interplay between boundary conditions and Wilson's mass in Dirac-like Hamiltonians}, Phys. Rev. B \textbf{100}, 205111 (2019).
\bibitem{Bee24} C. W. J. Beenakker, \textit{Chiral edge mode for single-cone Dirac fermions}, Phys. Rev. B \textbf{110}, 165421 (2024). The angles $\theta_\pm$ of that paper are related to the present $\theta_1,\theta_2$ by $\theta_-=\pi-\theta_1$, $\theta_+=-\theta_2$.
\bibitem{note1} In a graphene nanoribbon Fig.\ \ref{fig_masszigzag}d corresponds to zigzag edges on both sides, while Fig.\ \ref{fig_masszigzag}c corresponds to a zigzag edge on one side and a bearded edge on the other side. On the surface of a 3D topological insulator Figs.\ \ref{fig_masszigzag}c and \ref{fig_masszigzag}d corresponds to a magnetization parallel to the boundary and pointing, respectively, in opposite directions or in the same direction on the two sides of the channel.
\bibitem{Fuj96} M. Fujita, K. Wakabayashi, K. Nakada, and K. Kusakabe, \textit{Peculiar localized state at zigzag graphite edge}, J. Phys. Soc. Japan \textbf{65}, 1920 (1996).
\bibitem{Wak01} K. Wakabayashi, \textit{Electronic transport properties of nanographite ribbon junctions}, Phys. Rev. B \textbf{64}, 125428 (2001).
\bibitem{Koh07} M. Kohmoto and Y. Hasegawa, \textit{Zero modes and the edge states of the honeycomb lattice}, Phys. Rev. B \textbf{76}, 205402 (2007).
\bibitem{Vol09} V. A. Volkov and I. V. Zagorodnev, \textit{Electron states near graphene edge}, J. Phys. Conf. Ser. \textbf{193}, 012113 (2009).
\bibitem{Tka09} G. Tkachova and M. Hentschel, \textit{Spin-orbit coupling, edge states and quantum spin Hall criticality due to Dirac fermion confinement: the case study of graphene}, Eur. Phys. J. B \textbf{69}, 499 (2009).
\bibitem{Ost11} J. A. M. van Ostaay, A. R. Akhmerov, C. W. J. Beenakker, and M. Wimmer, \textit{Dirac boundary condition at the reconstructed zigzag edge of graphene}, Phys. Rev. B \textbf{84}, 195434 (2011).
\bibitem{Sus77} L. Susskind, \textit{Lattice fermions}, Phys. Rev. D \textbf{16}, 3031 (1977).
\bibitem{Her12} A. R. Hern\'{a}ndez and C. H. Lewenkopf, \textit{Finite-difference method for transport of two-dimensional massless Dirac fermions in a ribbon geometry}, Phys. Rev. B \textbf{86}, 155439 (2012). The Hamiltonian \eqref{Hstaggered} is equivalent to the transfer matrix given in that paper. (Note a typo: the factor $1/2\Delta$ in their equations 8 and 9 should apply to the entire right-hand-side, not just to the first term.)
\bibitem{Don23} A. Don\'{i}s Vela, G. Lemut, J. Tworzyd{\l}o, and C. W. J. Beenakker, Annals of Physics \textbf{456}, 169208 (2023).

\end{thebibliography}
\end{document}